\documentclass{aa}
\usepackage{graphics}
\usepackage{graphics}

\def\spose#1{\hbox to 0pt{#1\hss}}
\def\gsim{\mathrel{\spose{\lower 3pt\hbox{$\mathchar"218$}}
          \raise 2.0pt\hbox{$\mathchar"13E$}}}
\def\lsim{\mathrel{\spose{\lower 3pt\hbox{$\mathchar"218$}}
          \raise 2.0pt\hbox{$\mathchar"13C$}}}

\begin{document}

\title{The rise of the afterglow in  GRB 050820a}

\author{F. Genet, F. Daigne \& R. Mochkovitch}

\institute{Institut d'Astrophysique de Paris - UMR 7095 CNRS et Universit\'e
Pierre et Marie Curie, \\
98 bis, boulevard Arago, 75014 Paris, France\\ 
\tt e-mail: genet@iap.fr}
\titlerunning{The rise of the afterglow in  GRB 050820a}
\abstract
{The early optical afterglow of GRB 050820a recorded by the RAPTOR
telescope shows both a contribution from the prompt emission and 
the initial rise of the afterglow.}
{It is therefore well-suited for the study of the dynamical evolution
of the GRB ejecta when it first undergoes the decelerating
effect of the environment. This is a complex phase where the internal,
reverse, and forward shocks can all be present simultaneously.}
{We have developed a simplified model that can follow these different
shocks in an approximate, but self-consistent way. It is applied to the 
case of GRB 050820a to obtain the prompt and afterglow light curves.}
{We show that the rise of the afterglow during the course of the 
prompt emission has some important consequences. The reverse shock 
propagates back into the ejecta before internal shocks are completed, 
which affects the shape of the gamma-ray profile.}
{We get the best results when the external medium has a uniform 
density, but obtaining 
a simultaneous fit of the prompt and afterglow emission
is not easy. We discuss a few possibilities that could 
help to improve this situation.}

\authorrunning{F. Genet et al}
\titlerunning{The rise of the afterglow in  GRB 050820a}

\maketitle

\section{Introduction}
Just a few GRBs have been captured in the optical while the gamma-ray prompt
emission was still active. 
During the pre-SWIFT era, the only known example was GRB 990123. Its optical
signal was not correlated
to the gamma rays (Akerlof et al. 1999) and instead showed a power-law 
decline ($F\sim t^{-2}$), which was interpreted as a contribution from the 
reverse shock 
(Sari \& Piran 1999). In the past two years, thanks to the early and accurate
localizations by SWIFT and to the progress in ground-based robotic telescopes,
a few other cases have been discovered. While in GRB 050401 (Rykoff et al, 2006)
and GRB 051111 (Butler et al, 2006) only two data points were 
recorded during the gamma-ray emission, 
the TAROT observatory allowed continuous 
monitoring of GRB 060111b for more than 20 s during the prompt phase
(Klotz et al. 2006). In none of these bursts does the optical emission
appear correlated to the gamma rays, and GRB 060111b showed an initial
power-law decay of slope $\alpha=-2.38\pm 0.11$, very similar to the behavior
of GRB 990123. 
Conversely, the optical and gamma-ray light curves were
correlated in GRB 041219a (Vestrand et al. 2005), indicating that, at least in some bursts, 
the prompt component could also be detected in the
visible.

The case of GRB 050820a is especially interesting because, in addition to a prompt
component, the visible light curve exhibited a sharp rise at about 250 s, which
probably marks the onset of the afterglow. RAPTOR observations 
started 18 s after the BAT trigger that was caused by a faint precursor
200 s before the main pulses in the burst profile. Vestrand et al. (2006)
interpret the RAPTOR observations with a phenomenological function
describing the rise of the afterglow with an additional component 
proportional to the prompt KONUS gamma-ray profile. We use here a more
detailed description of the burst dynamical evolution to see if it can
reproduce the observations both in the gamma and visible ranges. 
\section{The burst model}
We have developed an approach that allows us to simultaneously follow 
the internal, reverse, and forward shocks in GRBs. This is crucial for
representing the early afterglow where these three kinds of shocks 
can coexist and contribute to the observed emission.  
Internal shocks are treated by using a large number of discrete shells to represent
the relativistic flow emerging from the central engine
(Daigne \& Mochkovitch 1998). Fast shells catch up and collide
with slower ones, the dissipated energy being shared between baryons,
electrons (fraction $\epsilon_e$), and the magnetic field 
(fraction $\epsilon_B$). Electrons then radiate by the synchrotron
process at a characteristic energy $E_{\rm syn}$, which depends on 
the assumed values of
$\epsilon_e$ and $\epsilon_B$. The spectrum is a broken power law of respective
(photon) index $\alpha$ and $\beta$ below and beyond $E_{\rm syn}$.
We have adopted $\beta=-2.25$ (Preece et al. 2000) and considered different
possible va\-lues for $\alpha$ between $-3/2$ and $-2/3$ (see Sect.3.2).

The interaction with the environment is implemented by
consi\-de\-ring the contact discontinuity that separates the ejecta and the shocked
external medium.
In our simple description, it is represented
by two shells moving at the same Lorentz factor $\Gamma$.
The first one corresponds to the part of the
ejecta already crossed by the reverse shock (of mass $M_{\rm ej}$)
and the second
to the shocked external medium (of mass $M_{\rm ex})$.  
Two processes affect this two-shell structure 
at the contact discontinuity: it collides either with shells of the
external medium at rest or with rapid shells from the relativistic ejecta
catching up. 
This represents both the forward and reverse shocks in our simplified 
picture (Genet, Daigne \& Mochkovitch 2006a).

The interaction with
the external medium is discretized by assuming that
a collision occurs each time
the contact discontinuity has travelled from a radius $R$
to a radius $R^{\,\prime}$, so that the swept-up mass is
\begin{equation}
m_{\rm ex}=\int_R^{R^{\,\prime}} 4\pi r^2 \rho(r) dr=q\,{M\over
\Gamma}
\end{equation}
where $M=M_{\rm ej}+M_{\rm ex}$, and $\rho(r)$ is the density of the external
medium. We have taken
$q=10^{-2}$, which represents a good compromise between numerical 
accuracy and computing time. 
\section{The prompt emission of GRB 050820a}
\subsection{Prompt gamma-ray emission}
To compute the prompt emission of GRB 050820a, we start from
an initial distribution of the Lorentz factor in the flow ejected 
by the central source, which  can lead to the observed gamma-ray 
profile. Such a distribution is shown in Fig.1. It is made of 
several episodes of wind production for a total duration of about 200 s
in the source rest frame. In each episode 
``slow" material ($\Gamma=100$) is emitted first and then followed by some
more rapid one ($\Gamma=400$) with a transition of the form
\begin{equation}
\Gamma(t)=250-150\,{\rm cos}\left[\pi\left({t-t_0\over t_{\rm m}-t_0}
\right)\right]
\end{equation}
where $t_0$ is the starting time of the episode and $t_{\rm m}$ 
the time when $\Gamma(t)$ reaches its maximum value of 400. 
Such a cosine form has been used in our previous works (Daigne
\& Mochkovitch 1998, 2000) and provides a smooth transition between
the rapid and slower parts of the flow.
The kinetic
energy injected in the different episodes is fixed to reproduce the 
intensity of the successive spikes in the gamma-ray profile.
To account for the high isotropic gamma-ray energy of GRB 050820a
($E_{\gamma}^{\rm iso}\sim 8\,10^{53}$ erg, Golenetskii et al. 2005), 
we had to inject
an even higher kinetic energy $E_{\rm K}^{\rm iso}=1.8\;10^{55}$ erg 
into the flow
since the global efficiency $f$ of the conversion process is low.
We have $f=f_{\rm diss}\times \epsilon_e\lsim 5\%$, 
where $f_{\rm diss}\lsim 
15\%$ is the efficiency for dissipation by internal shocks. 
We have assumed a high $\epsilon_e=0.33$ to still have
a reasonable total efficiency, since only the fraction of the
energy transferred to electrons is finally radiated.
We compare the resulting synthetic profile  (obtained with
a low energy index $\alpha=-1$) to the KONUS 
light curve in Fig.2, neglecting at this stage the effect of the external medium.
The agreement is satisfactory, as the
main objective of this work is not to reproduce the temporal 
behavior of GRB 050820a accurately but to study the rise of the afterglow
in a complex burst.
\begin{figure*}{}
\begin{center}
\begin{tabular}{cc}
\resizebox{7cm}{7cm}{\includegraphics{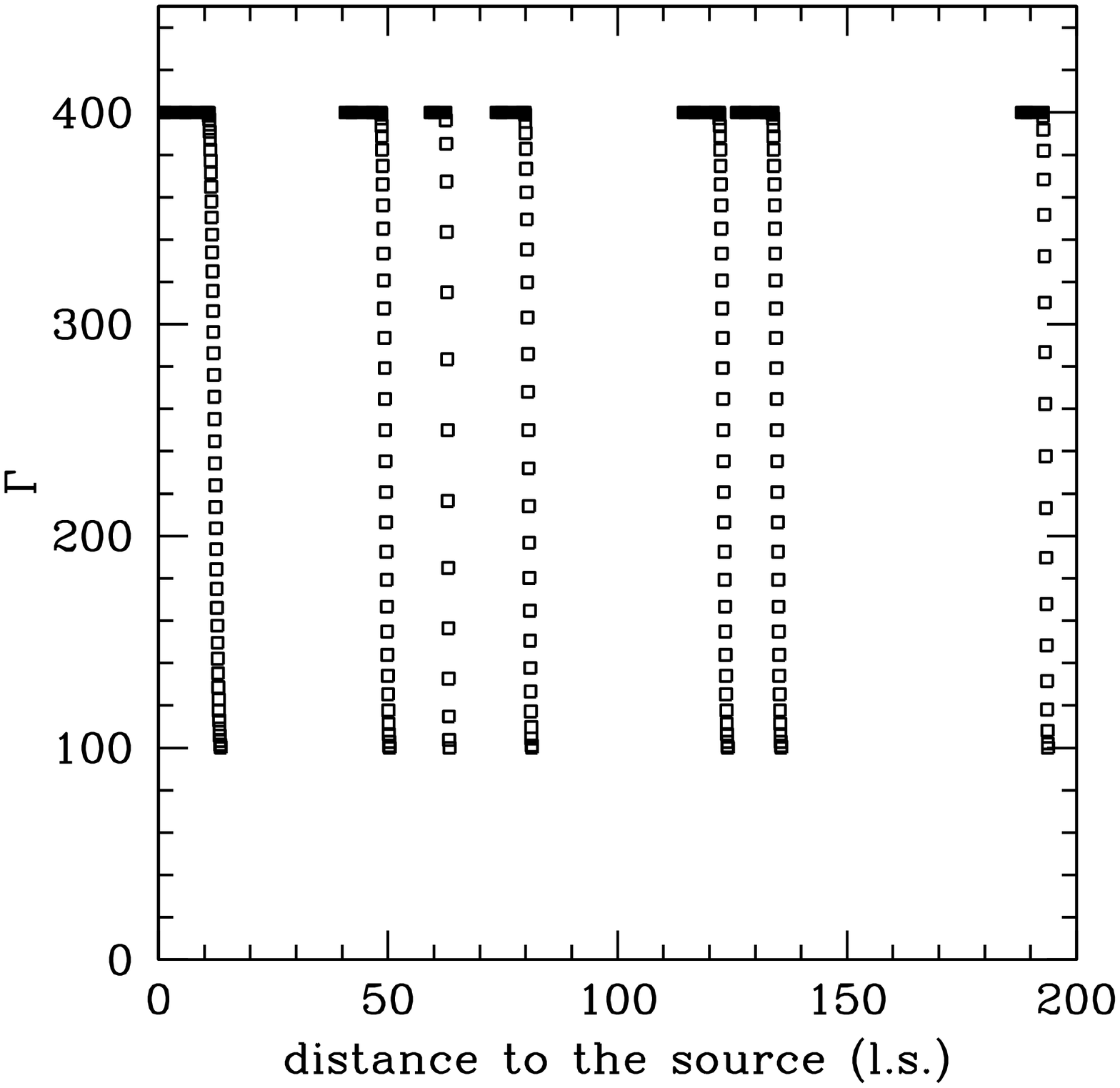}}&
\resizebox{7cm}{7cm}{\includegraphics{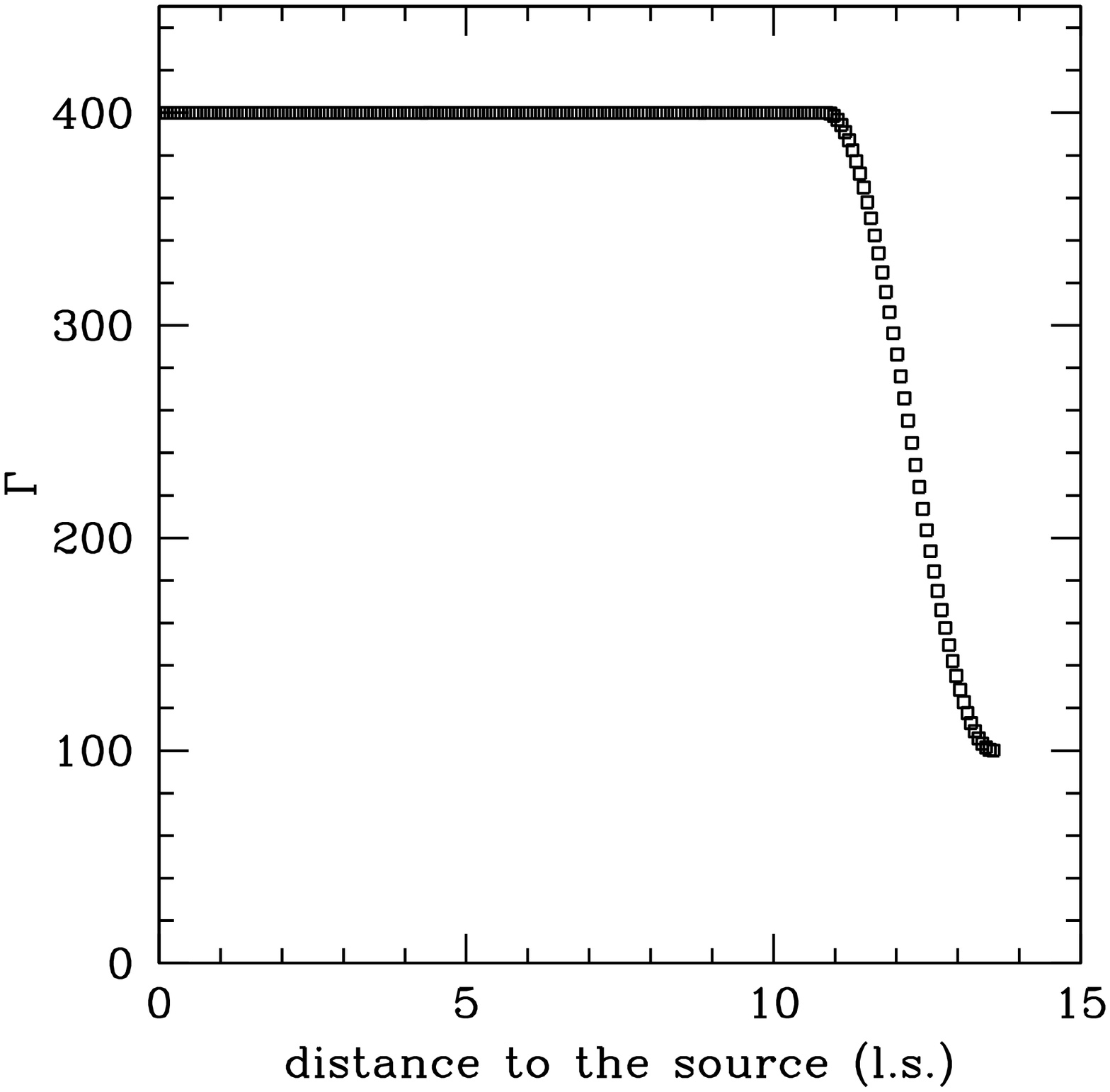}}\\
\end{tabular}
\end{center}
\caption{Left: initial distribution of the Lorentz factor in the relativistic
flow. The $\Gamma$ value of each shell is given as a function of the distance to the 
source in light seconds. Right: zoom on the last episode of wind production
showing the transition between the slow and fast material. }
\end{figure*}
\subsection{Prompt optical emission}
\begin{figure*}{}
\begin{center}
\begin{tabular}{cc}
\resizebox{7cm}{7cm}{\includegraphics{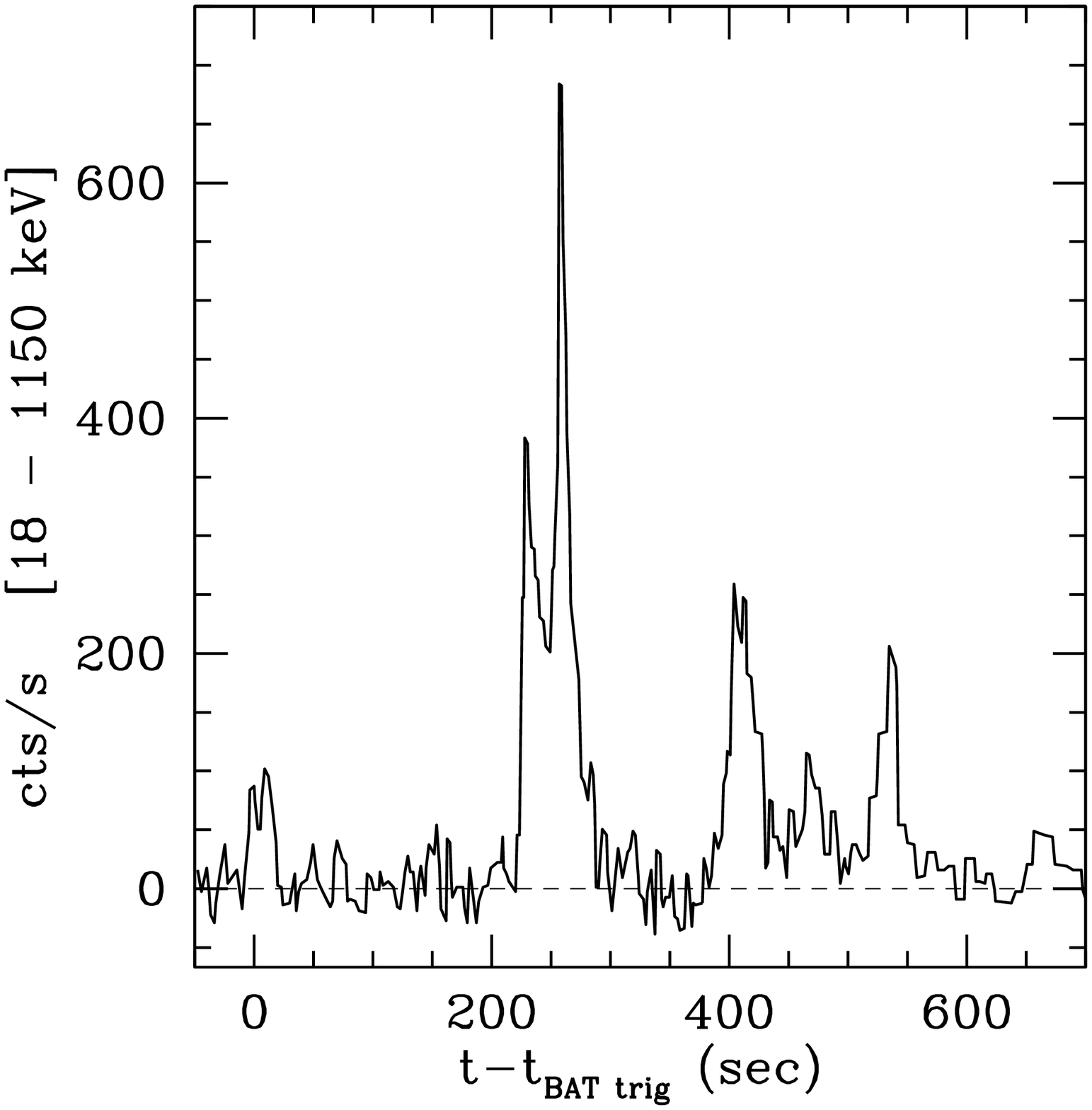}}&
\resizebox{7cm}{7.cm}{\includegraphics{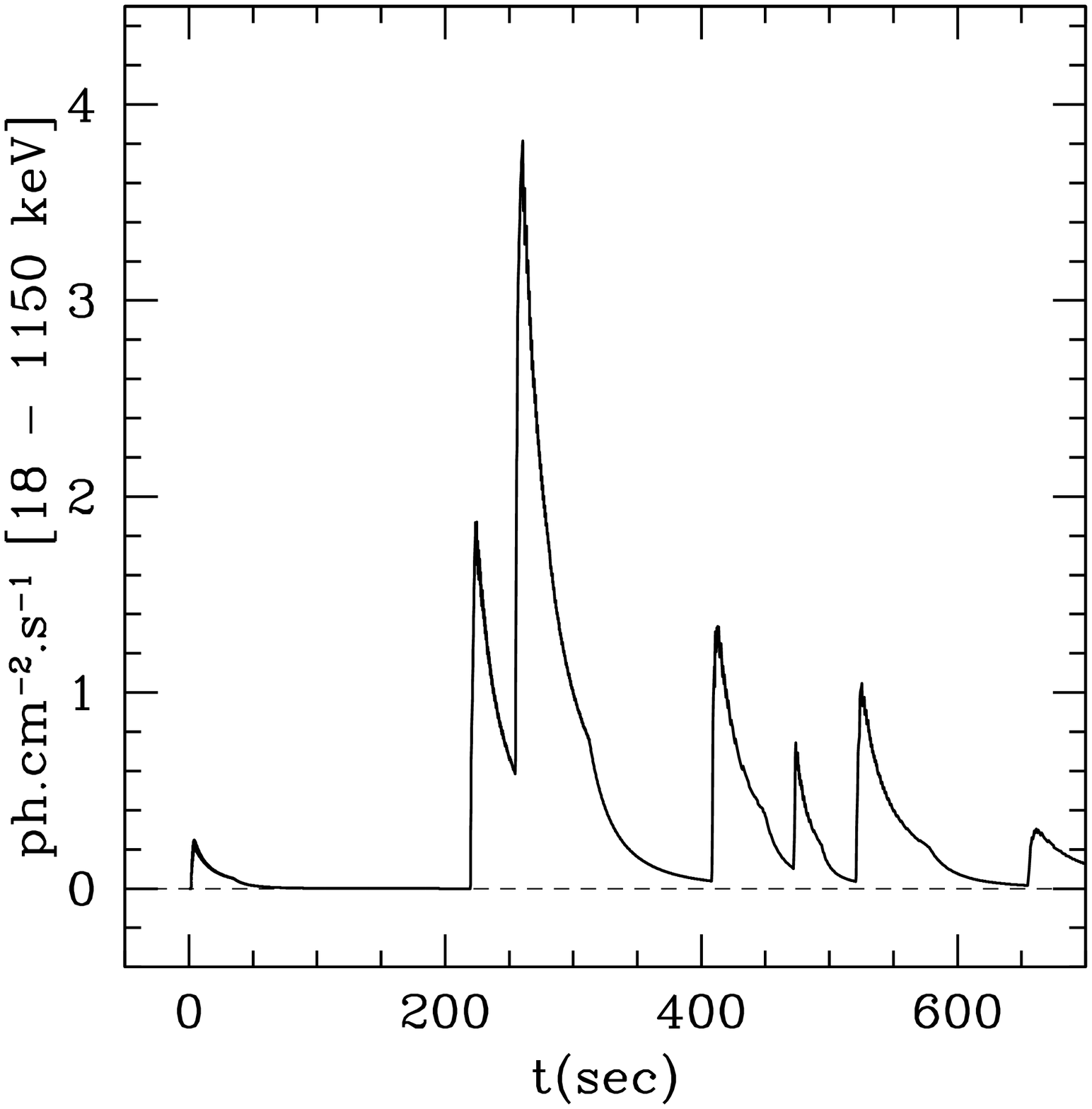}}\\
\end{tabular}
\end{center}
\caption{Comparison of the KONUS light curve (left) to our synthetic 
profile (right) in the same energy band: 18 - 1150 keV. } 
\end{figure*}
The RAPTOR observations  show a contribution from the prompt
emission in the optical that becomes blended with the afterglow after 
about 200 s. 
The prompt emission of GRBs must take place in the fast cooling
regime to gua\-ran\-tee that the energy dissipated by internal shocks is
efficiently radiated. The expected emission spectrum
should then have a spectral index
of $\alpha=-3/2$ for $\nu_c<\nu<\nu_m$ and $\alpha=-2/3$ for $\nu<\nu_c$,
where $\nu_m$ is the characte\-ris\-tic synchrotron frequency 
and $\nu_c$ the cooling frequency that typically lies
in the optical/UV range (Sari, Piran \& Narayan 1998).
However such a spectrum contradicts the majority of 
observed GRB spectra (Ghisellini, Celotti \& Lazzati, 2000) 
where the typical low-energy photon spectral 
index $\alpha$
is closer to $-1$ than to $-1.5$ (Preece et al, 2000). 
A more detailed description of the emission processes
would therefore be necessary to obtain the prompt optical flux
in a fully reliable way. Here we have simply adopted a single, 
averaged low-energy spectral index $\langle\alpha\rangle$ from the gamma
to the optical energy ranges.  The resulting prompt optical flux is
then very sensitive to the value of $\langle\alpha\rangle$
as shown in the left panel of Fig.3 where the R magnitude light curve has been represented
for $\langle\alpha\rangle=-2/3$, $-1$ and $-3/2$.
\begin{figure*}{}
\begin{center}
\begin{tabular}{cc}
\resizebox{7cm}{7cm}{\includegraphics{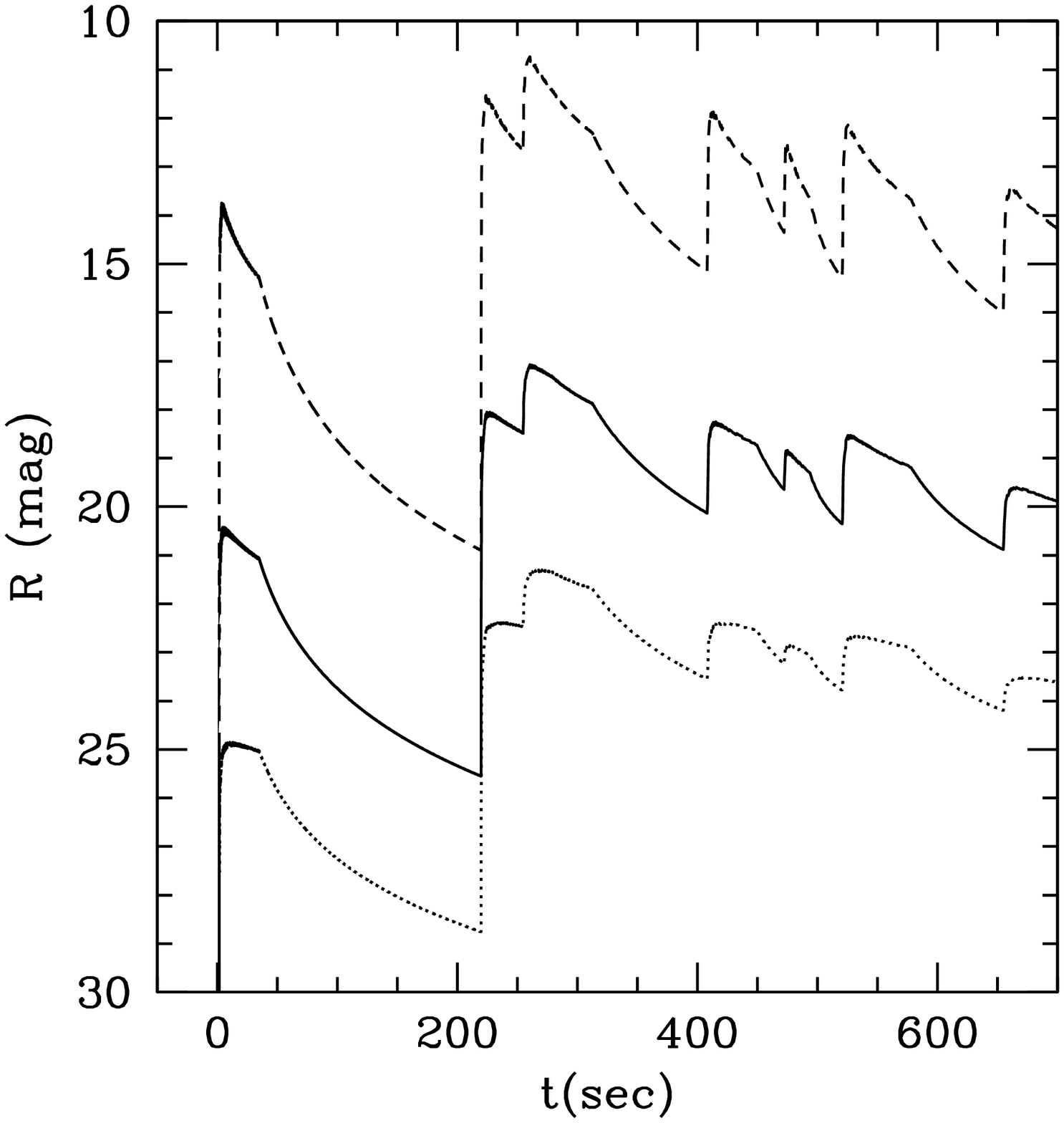}}&
\resizebox{7cm}{7cm}{\includegraphics{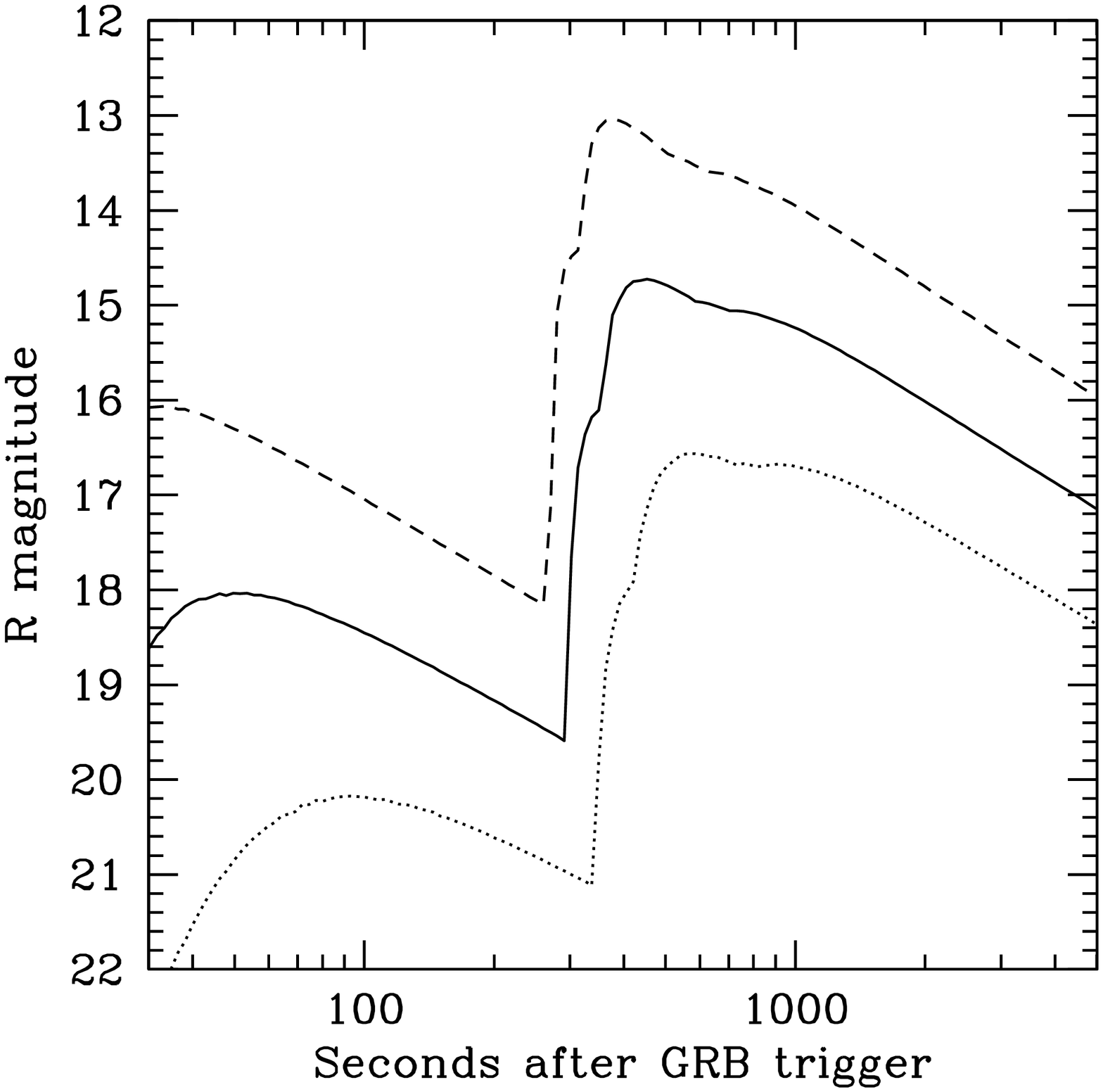}}\\
\end{tabular}
\end{center}
\caption{Left: prompt light curves in the R band for 
$\langle\alpha\rangle=-2/3$ (dotted line), $-1$ (full line), and $-1.5$
(dashed line); Right: afterglow light curves for 
$\epsilon_e=3\,10^{-3}$, $\epsilon_B=10^{-5}$,
$p=2.5$,
and different densities in the
external medium, respectively $n=0.3$ (dotted line), 3 (full line), and 
30 cm$^{-3}$ (dashed line).}
\end{figure*}
When $\langle\alpha\rangle$ increases from $-3/2$ to $-2/3$, the R flux 
decreases by about 10 magnitudes! 
Since the prompt contribution to the RAPTOR light curve 
peaks at $R\sim 15$, it indicates a value of  
$\langle\alpha\rangle\sim -1.15$.
We also find that the profile shape evolves, becoming
less spiky at larger $\langle\alpha\rangle$.
This can be understood since the flux in the visible is proportional
to $\left({E_R\over E_{\rm syn}}\right)^{\langle\alpha\rangle+1}$, where
$E_R\sim 1$ eV is a typical energy for the R band. Comparing the 
fluxes for two different values of $\langle\alpha\rangle$ yields
\begin{equation}
{F_R^{\langle\alpha_1\rangle}\over F_R^{\langle\alpha_2\rangle}}=
\left({E_R\over E_{\rm syn}}\right)^{\langle\alpha_1\rangle-
\langle\alpha_2\rangle}\ .
\end{equation} 
If $E_{\rm syn}$ stayed constant during burst evolution, the light curves for
different $\langle\alpha\rangle$ would be simply proportional. 
However, since
$E_{\rm syn}$ is correlated to intensity, the flux ratio varies: 
for $\langle\alpha_1\rangle<
\langle\alpha_2\rangle$, it 
increases
at intensity peaks.  
\section{Afterglow calculation}
\begin{figure}
\begin{center}
\resizebox{7cm}{7cm}{\includegraphics{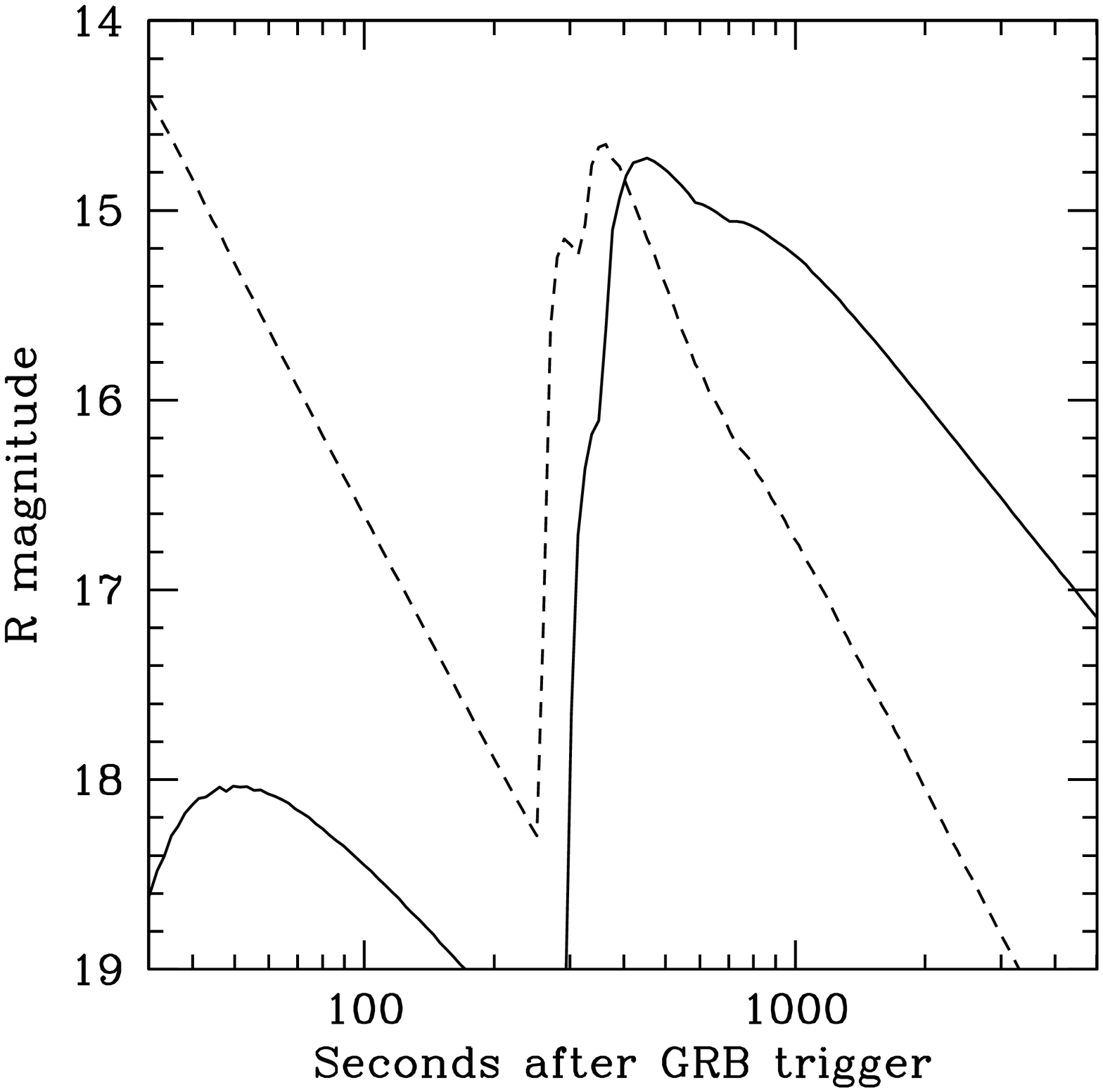}}
\end{center}
\caption{Comparison of the afterglow light curve for a wind 
with $A_*=0.1$ (dashed line) and a uniform medium with $n=3$ cm$^{-3}$ (full line).}
\end{figure}
We first obtained afterglow light curves for a uniform external medium. 
The general shape of the afterglow consists of an initial weak bump produced
when the material responsible for the gamma-ray precursor hits the ISM. It is
followed by a sharp rise leading to the second bright bump when the
material making the main peaks of the profile catches up with the 
forward shock. 
The right panel of Fig.3 shows the evolution of the afterglow when the 
external density
$n$ is varied. Increasing $n$ naturally leads to a brighter 
afterglow with an earlier rise of the second bump. 
Once the density has been fixed by the time of the onset of
the second bump, the microscopic parameters $\epsilon_e$, $\epsilon_B$,
and $p$ can be adjusted to get the correct intensity and decay slope
after maximum. A good compromise appears to be 
$n=3$ cm$^{-3}$, $\epsilon_e=3\,10^{-3}$, $\epsilon_B=10^{-5}$, and
$p=2.5$.

We then considered a stellar wind environment, which should be expected 
if long bursts are produced by explo\-ding WR stars. An example of the 
resulting afterglow is given in Fig.4. It can be seen that 
the first bump is now much too bright since the first part of the ejecta
encounters the densest part of the wind, close to the 
star. The RAPTOR observations (where the first bump is 3 magnitudes
fainter than the second) then clearly favor a uniform environment
except if some special circumstances (varying microphysics 
parameters, pair loading and pre-acceleration of the external medium) can strongly reduce
the first bump's contribution.  
\section{Effect of the environment on the burst prompt emission}
The afterglow light curve obtained with $n=3$ cm$^{-3}$ fits 
the RAPTOR data reasonably well, as shown in Fig.5 where we also represent 
the prompt emission component for $\langle\alpha\rangle\sim -1.15$. 
However the rise of the afterglow 
at $t\sim 250$ s, i.e. du\-ring the prompt phase, has 
some important consequences. The reverse shock
resulting from the early deceleration of the ejecta affects the 
distribution of the Lorentz factor well before internal shocks are completed.
This is turn modifies the gamma-ray profile as shown in Fig.6, where the
profile with $n=3$ cm$^{-3}$ is compared to the $n\rightarrow 0$ case
already shown in Fig.2. After 300 s, new pulses are present, the 
photon flux is larger,   
and the similarity with
GRB 050820a is partially lost. Reducing the density of the external
medium tends to better preserve the burst profile but also
shifts the rise of the afterglow to later times (see Fig.3).
\begin{figure}
\begin{center}
\resizebox{7cm}{7cm}{\includegraphics{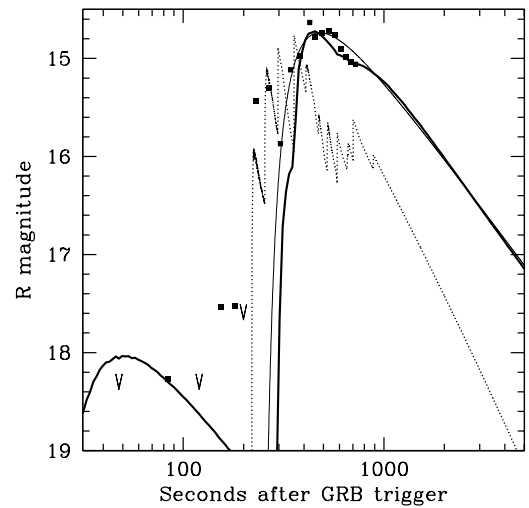}}
\end{center}
\caption{Results of our model compared to the RAPTOR observations. The thick
full line is our synthetic afterglow light curve (external 
shock contribution)
 and the dotted line
is the prompt optical component calculated with
$\langle\alpha\rangle\sim -1.15$ (internal shock contribution). RAPTOR 
data (Vestrand et al. 2006) 
is shown as squares or V
signs (upper limits). The thin full line is a phenomenological
function used by Vestrand et al. to represent the afterglow component.}
\end{figure}

\begin{figure}
\begin{center}
\resizebox{7cm}{7cm}{\includegraphics{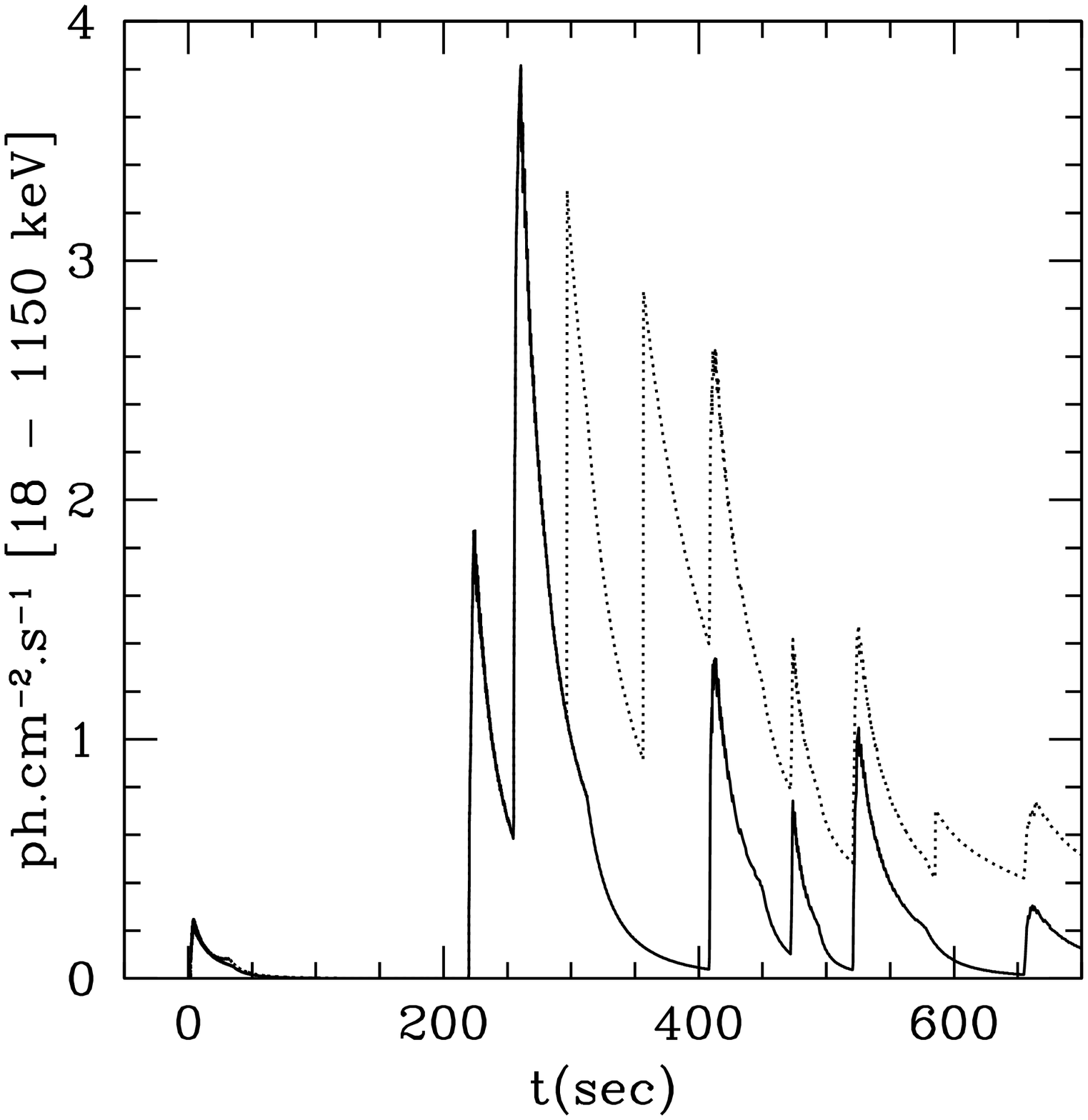}}
\end{center}
\caption{Predicted gamma-ray profile when the effect of the external
medium is included (for $n=3$ cm$^{-3}$, dotted line) compared to the 
original profile with no external medium (full line).}
\end{figure}
\section{Discussion}
The simplified model we have developed neglects many important aspects of 
the hydrodynamical evolution of the relativistic outflow
produced by the burst central engine (e.g. pressure waves, radial structure 
of flow). However, within these approximations, it allows us 
to study the complete evolution with one single, self-consistent 
calculation including both the phase of internal shocks and the deceleration
by the external medium. It is therefore ideally suited for
GRB 050820a where the afterglow starts to rise before the end of internal 
shocks.
 
We have shown that obtaining a simultaneous 
fit for both the prompt and afterglow emission of GRB 050820a is not
easy  because, due to the early deceleration of the flow, the reverse shock 
gets mixed with internal shocks, which affects the burst profile in a 
complicated way. 
We are therefore currently examining a few possibilities that could improve
the situation:
\vskip 0.3cm\noindent (i) The most obvious solution would be 
to start with an initial distribution of the Lorentz factor
different from the one shown in Fig.1 so that, after it had been perturbed by 
the reverse shock, it would finally lead to the observed gamma-ray profile. Preliminary
calculations indicate that it should be possible to recover the main
spikes in the profile, 
but also difficult to avoid a few additional weak pulses partially
overlapping with them.  
\vskip0.3cm\noindent
(ii) Another possibility could be to include in the calculation the pair-loading process resulting from the
gamma-ray flash that preaccelerates the circumstellar medium (Madau \&
Thomson, 2000; Beloborodov, 2002). The low-density cavity that forms
around the source will reduce the effect of the reverse shock. It will 
however also
delay the rise of the afterglow by an amount that should remain compatible 
with the observations;
\vskip0.3cm\noindent
(iii) One could finally consider the electromagnetic model (EMM) proposed by 
Lyutikov \& Blandford 
(2003) rather than the standard model. The EMM has no reverse shock, but
its physics is also more complex and uncertain so that
the comparison with  observations is not straightforward. 
Using a simple model to compute GRB afterglows in the context of the EMM
(Genet, Daigne \& Mochkovitch 2006b), we have found that good fits 
of the GRB 050820a afterglow can be obtained if electromagnetic energy
is released in two steps: a weak precursor followed by the main event.
However, we are unable to calculate the related prompt optical and
gamma-ray emissions that, in the EMM, result from magnetic 
reconnection processes. It is  
therefore impossible to check the overall consistency of the model.
\section{Conclusion}
We have developed an approximate method that can follow the complex
dynamical evolution of GRB ejecta du\-ring the early afterglow phase
when the internal, reverse, and forward shocks can be simultaneously
active and contribute to the observed radiation.

It has been applied to GRB 050820a, for which RAPTOR observations show
both a prompt emission component and the onset of the afterglow. We have found
that the rise of the afterglow during the prompt phase implies that the
reverse shock is active well before internal shocks are completed.
This makes a simultaneous fit of the prompt gamma-ray emission and 
optical afterglow rather challenging, the best results being however
obtained for a uniform external medium.
We have briefly discussed possible ways to improve the situation
in the context of the standard internal/external shock 
model but also considered the EMM as an alternative.

As more bursts are captured in the optical du\-ring the prompt phase
and very early afterglow, 
it should become possible to study in more detail this complex period
where the burst energy starts to be transferred to the surrounding medium.
In X-rays it is often characterized by a shallow decline 
that also challenges simple models. Multiwavelength observations
showing the rise of the afterglow
will then be of prime interest to see if it can still be explained
by the standard internal/external shock model 
or if changes in the current paradigm will be necessary.

\end{document}